\documentclass[showpacs,amsfonts,amssymb,amsmath,preprint,floatfix,superscriptaddress]{revtex4-1}
\usepackage{graphicx}
\usepackage{color}

\begin{document}

\title{Langevin dynamics for ramified structures}
\author{Vicen\c{c} M\'{e}ndez}
\affiliation{Grup de F{\'i}sica Estad\'{i}stica.  
Departament de F{\'i}sica. Facultat de Ci{\`e}ncies. 
Edifici Cc. Universitat Aut\`{o}noma de Barcelona, 08193 
Bellaterra (Barcelona) Spain} 

\author{Alexander Iomin} 
\affiliation{Department of Physics, Technion, Haifa, 32000, Israel}

\author{Werner Horsthemke}
\affiliation{Department of Chemistry, 
Southern Methodist University,
Dallas, Texas 75275-0314, USA}

\author{Daniel Campos} 
\affiliation{Grup de F{\'i}sica Estad\'{i}stica.  
Departament de F{\'i}sica. Facultat de Ci{\`e}ncies. 
Edifici Cc. Universitat Aut\`{o}noma de Barcelona, 08193 
Bellaterra (Barcelona) Spain}

\date{\today}

\begin{abstract}
We propose a generalized Langevin formalism to describe
transport in combs and similar ramified structures. Our
approach consists of a Langevin equation without drift
for the motion along the backbone. The motion along the 
secondary branches may be described either by a
Langevin equation or by other types of random processes. 
The mean square displacement (MSD) along the backbone
characterizes the transport through the ramified structure. 
We derive a general analytical expression
for this observable in terms of the probability distribution
function of the motion along the secondary branches. We apply
our result to various types of motion along the
secondary branches of finite or infinite length, 
such as subdiffusion, superdiffusion, and
Langevin dynamics with colored
Gaussian noise and with non-Gaussian white noise. 
Monte Carlo simulations show excellent agreement
with the analytical results.
The MSD for the case of Gaussian noise  
is shown to be independent of the noise color.
We conclude
by generalizing our analytical 
expression for the MSD
to the case where each secondary branch is $n$ dimensional. 

\end{abstract}

\maketitle

\section{Introduction}\label{sec:intro}
Various phenomena in physics,
biology, geology, and other fields
involve the transport or motion
of particles, microorganisms, and 
fluids in ramified structures. 
Examples range from fluid flow 
through porous media to oil recovery, 
respiration, 
and blood circulation. Ramified 
structures like river networks \cite{RIRi97} 
represent examples of ecological 
corridors, which have significant
implications in epidemics
\cite{BeCaGaRoRi10} or diversity 
patterns \cite{MuBeLyFaRi08}, among other. 
Ramified structures have also attracted
the attention of physicists 
because 
the transport of particles across them 
displays anomalous diffusion \cite{FoBuCeVu13}. 

The simplest models of these various 
types of natural structures, which belong 
to the category of
loopless graphs, are
the comb 
model and the Peano network, two ramified 
structures that have been applied, for example, to 
explain biological invasion through 
river networks \cite{MeFeHo10}. 
Comb structures consist of 
a principal branch, the backbone,
which is a one-dimensional 
lattice with spacing $a$,  
and identical secondary
branches, the teeth,
that
cross the backbone perpendicularly.
We identify the direction of the
backbone with
the $x$-axis, while
the secondary branches lie
parallel to the $y$-axis.
Nodes on the backbone have the 
coordinates $(ia,0)$, with
$i=0,\pm 1,\pm 2, \dotsc$, while
nodes on the teeth have
coordinates $(ia,ja)$, with
$j=0,\pm 1,\pm 2, \dotsc$ and $i$ fixed.

The comb model was originally introduced to 
understand anomalous diffusion 
in percolating clusters \cite{WeHa86,WhBa84,ArBa91}.
If particles undergo a simple
random walk on the comb structure,
the secondary branches act like
traps in which the particle stays 
for some random time before continuing its 
random motion along the backbone.
This results in a mean square displacement
(MSD) $\left\langle x^{2}(t)\right\rangle \sim
\sqrt{t}$, i.e., subdiffusive behavior
along the backbone.
Nowadays, comb-like models are widely 
used to describe different experimental 
applications, such as anomalous transport 
along spiny dendrites \cite{MeIo13,IoMe13,FeMe08} and  
dendritic polymers \cite{Fr05f}, to mention 
just a few.

In the continuum limit, transport on a comb
can be described by an anisotropic
diffusion equation,      
\begin{equation}
\label{eq:combDE}
\frac{\partial P(x,y,t)}{\partial t}=
[C(y)]^2D_{x}\frac{\partial^{2}P(x,y,t)}{\partial x^{2}}
+D_{y}\frac{\partial^{2} P(x,y,t)}{\partial y^{2}}.
\end{equation}
This diffusion equation is equivalent
to the system of Langevin equations
\begin{subequations}
	\label{eq:leGWN}
	\begin{align}
	\frac{dX}{dt}&=C(Y)\xi_{x}(t),\label{eq:leq1GWN}\\
	\frac{dY}{dt}&=\xi_{y}(t),\label{eq:le2GWN}
	\end{align}
\end{subequations}
where $\xi_{x}(t)$ and $\xi_{y}(t)$ are two uncorrelated
Gaussian white noises with
\begin{subequations}
	\label{eq:noisechar}
	\begin{align}
	\left\langle\xi_{x}(t)\right\rangle&=\left\langle\xi_{y}(t)
	\right\rangle=0,\\
	\left\langle \xi_{x}(t)\xi_{x}(t')\right\rangle &
	=2D_{x}\delta(t-t'),\\
	\left\langle \xi_{y}(t)\xi_{y}(t')\right\rangle &
	=2D_{y}\delta(t-t'),\\
	\left\langle \xi_{x}(t)\xi_{y}(t')\right\rangle &
	=0.
	\end{align}
\end{subequations}
Here $\langle \cdot\rangle$ denotes averaging over the 
noises. Eq. (\ref{eq:combDE}) can be obtained assuming both Ito and Stratonovich interpretations since the specific form of the Langevin equations (\ref{eq:leGWN}) yields to the same Wong-Zakai terms.

The coefficient $C(y)$ in Eq.\ (\ref{eq:combDE}) 
introduces a heterogeneity that couples the motion in 
both directions. In most works about transport on 
combs \cite{ArBa91,Ar99,BaIo04} this coefficient 
is taken to be $\left[C(y) \right]^2=\delta(y)$, 
a Dirac delta function,
which means that the
teeth cross the backbone only at $y=0$. 
The system of
equations \eqref{eq:leGWN} has
also been applied to certain problems
in biochemical
kinetics \cite{BeNe09a,RiTaAlZoLe14}.

Our goal is to apply the Langevin equations
\eqref{eq:leGWN} to situations where the motion of particles does not correspond to simple
Brownian motion. In particular we will focus on the case where the driving noises along the teeth, i.e., in the $y$-direction are no longer Gaussian
white noises. In other words, we consider combs where the transport process
along the teeth can differ fundamentally from the transport process along
the backbone

The paper is organized as follows.
In Sec.~\ref{sec:LE} we introduce
our generalized Langevin description.
An exact analytical expression for
the MSD along the backbone is derived
in Sec.~\ref{sec:MSD}. We use that
result to investigate the effect
of subdiffusive and superdiffusive
motion along the teeth, 
motion driven by various types of noises,
as well as the effect of the geometry of
the teeth in Secs.~\ref{sec:teethinf}, \ref{sec:teethLang}, and \ref{sec:teethfin}.
We discuss our results in Sec.~\ref{sec:concl}.

\section{Langevin Equations}\label{sec:LE}

Consider first a ramified structure where 
the particle dynamics is governed
by the general Langevin equations
\begin{subequations}
\label{eq:genLE}
\begin{align}
\frac{dX}{dt}&=\beta_{x}C(Y)\xi_{x}(t),\label{eq:leq1}
\\
\frac{dY}{dt}&=\xi_{y}(t).\label{eq:le2}
\end{align}
\end{subequations}
Here $(X(t),Y(t))$ is a random process 
describing the position of the particle in a two-dimensional
space, and
$\beta_{x}$ is a 
positive parameter. 
The random driving forces
$\xi_{x}$ and $\xi_{y}$ are two 
external noises that drive the 
motion of the
particle along the $x$-direction, 
backbone or main direction,
and the $y$-direction, 
branches or secondary direction, respectively. 
The motion along the
$y$-direction is then independent of
the $x$ coordinate. 
The coupling of the motions along
the $x$ and $y$ directions is described
by $C(Y)$. In fact, we will
consider a more general system than 
Eqs.\ \eqref{eq:genLE}. The random
process $Y(t)$ does not have to be given
by the Langevin equation \eqref{eq:le2};
it can be any suitable random process
describing the motion in the $y$-direction,
as long as it is independent of $X(t)$.
In the following,
$\left\langle\cdot \right\rangle$ 
denotes averaging over one random variable, e.g. $X$,
and 
$\left\langle\left\langle\cdot \right\rangle
\right\rangle$ over all random variables involved,
e.g., $X$ and $Y$.
To determine the MSD 
we rewrite Eq.\ (\ref{eq:leq1})
in the form
\begin{equation}
\frac{d}{dt}(X^{2})=2\beta_{x}C[Y(t)]\xi_{x}(t)X(t).
\label{eq:x2}
\end{equation}
We integrate 
Eq.\ (\ref{eq:leq1}) with the initial
condition $X(0)=0$, substitute
the result into Eq.\ (\ref{eq:x2}), and
average over the noise $\xi_{x}(t)$ to find
\begin{equation}
\frac{d}{dt}\left\langle X^{2}(t)\right\rangle 
=2\beta_{x}^{2}C[Y(t)]\int_{0}^{t}C[Y(t')]
\left\langle \xi_{x}(t)\xi_{x}(t')\right\rangle dt'.
\label{eq:x22}
\end{equation}

In the following we assume in all
cases that the noise $\xi_{x}(t)$ 
driving
the motion along the 
backbone is
white, i.e., 
$\left\langle \xi_{x}(t)\xi_{x}(t')\right\rangle 
=2D_{x}\delta(t-t')$,
and we adopt the Stratonovich interpretation. 
We also consider for simplicity that both noises
$\xi_{x}$ and $\xi_{y}$ are uncorrelated. 
Then Eq.\ (\ref{eq:x22})
turns into
\begin{equation}
\frac{d}{dt}\left\langle X^{2}(t)\right\rangle 
=2D_{x}\beta_{x}^{2}          \left(C[Y(t)]\right)^{2}.
\label{eq:x23}
\end{equation}
Let $\mathcal{D}$ be the range of $Y(t)$.
Then averaging   
Eq.\ (\ref{eq:x23}) over $Y$, we obtain
\begin{equation}
\frac{d}{dt}\left\langle \left\langle 
X^{2}(t)\right\rangle \right\rangle
=2D_{x}\beta_{x}^{2}\int_{\mathcal{D}}(C[y])^2P_{Y}(y,t)dy,
\label{eq:x24}
\end{equation}
where
$P_{Y}(y,t)=\langle\delta (Y(t)-y)\rangle$.
Consequently, the MSD for transport through
the ramified structure is given by
\begin{equation}
\left\langle \left\langle 
X^{2}(t)\right\rangle \right\rangle
=2D_{x}\beta_{x}^{2}\int_{0}^{t}dt'\int_{\mathcal{D}}(C[y])^2P_{Y}(y,t')dy.
\label{eq:MSDx24}
\end{equation}

\section{The Mean Square Displacement}\label{sec:MSD}

We use the result \eqref{eq:MSDx24}
to assess the influence of various types of motion in the
$y$-direction on the transport through the structure.
The simplest case occurs if the structure
is actually not ramified at all, i.e., the
particles move in the $x$-$y$-plane. 
The dynamics of $X(t)$ and $Y(t)$ 
are independent, i.e., $C[Y(t)]=C=$ const. 
We obtain
from Eq.\ (\ref{eq:MSDx24}) 
\begin{equation}
\left\langle \left\langle X^{2}(t)\right\rangle\right\rangle
=2D_{x}\beta_{x}^{2}C^{2}t.
\end{equation}
In other words, the motion projected into
the $x$-axis corresponds to normal
diffusive behavior. This is the expected
result, since $X(t)$ does not depend
on $Y(t)$ and is driven by white noise.

More interesting behavior occurs for
a comb-like structure. To account for this case we consider that the coupling function can be written as
\begin{equation}
C[y]=\sqrt{\frac{\ell}{\pi (y^2+\ell ^2)}}.  
\label{Cy}
\end{equation}
Note that $C^2[y]$ is a regularization, or representation, of the Dirac delta function for $\ell \rightarrow 0$. So, invoking the fact that $C^2[y]\rightarrow \delta (y)$ for $\ell \rightarrow 0$, Eq.\ (\ref{eq:MSDx24}) 
reads
\begin{align}
\left\langle \left\langle X^{2}(t)
\right\rangle \right\rangle&=2\beta_{x}^{2}D_{x}
\int_{0}^{t}dt'
\int_{-\infty}^{\infty}P_{Y}(y,t')\delta (y)dy\nonumber\\
&=2\beta_{x}^{2}D_{x}\int_0 ^t dt'  P_{Y}(y=0,t')
\label{eq:r0}
\end{align}
or
in Laplace space
\begin{equation}
\left\langle \left\langle \hat{X}^{2}(s)
\right\rangle \right\rangle=
2\beta_{x}^{2}D_{x}\frac{\hat{P}_{Y}(y=0,s)}{s},
\label{eq:x2sk}
\end{equation}
where the hat symbol denotes the 
Laplace transform and $s$ is the Laplace variable.

Taking into account the inverse Fourier transform 
$P_Y(y,t)=(1/2\pi)\int_{-\infty}^{\infty} dk\exp(-iky)P_Y(k,t)$, it is easy to 
see that 
$\hat{P}_{Y}(y=0,s)=(1/2\pi)\int_{-\infty}^{\infty}dk\hat{P}_{Y}(k,s)$. 
Substituting this result into (\ref{eq:x2sk}),
we find that the MSD reads
\begin{equation}
\left\langle \left\langle \hat{X}^{2}(s)
\right\rangle \right\rangle=
\frac{\beta_{x}^{2}D_{x}}{\pi s}\int _{-\infty}^{\infty}dk\hat{P}_{Y}(k,s),
\label{eq:x2s}
\end{equation}
i.e., we can determine the MSD in Laplace space 
if we know the propagator, in Fourier-Laplace space, along the teeth.

\section{Secondary branches with infinite length}\label{sec:teethinf}
Note that our results for the MSD, Eqs.\ \eqref{eq:r0}
-- \eqref{eq:x2s}, are valid as long as
the movement 
along the backbone follows the Langevin dynamics 
given by Eq.\ (\ref{eq:leq1}).
The motion of the particles along the teeth need not 
be governed by the Langevin dynamics Eq.\ \eqref{eq:le2};
it can be any suitable random process. In this Section
we explore transport through the comb when the movement of particles along the teeth is anomalous, i.e., non-standard diffusion.

\subsection{Continuous-Time Random Walk}

We consider here the case where the motion
along the teeth can be described by a Continuous-Time Random Walk (CTRW).
The propagator in Fourier-Laplace space $\hat{P}_{Y}(k,s)$ 
is given, in general, by the Montroll-Weiss equation \cite{MoWe65}, and
we obtain from Eq.\ (\ref{eq:x2s}) 
\begin{equation}
\left\langle \left\langle \hat{X}^{2}(s)\right\rangle \right\rangle
=\frac{\beta_{x}^{2}D_{x}[1-\hat{\phi}(s)]}{\pi s^2}
\int_{-\infty}^{\infty}\frac{dk}{1-\lambda (k)\hat{\phi}(s)},
\label{eq:x2s2}
\end{equation}
where $\lambda (y)$ and $\phi(t)$ are the jump length and 
waiting time PDFs of the random motion along 
the branches, respectively.

Subdiffusive motion along the teeth occurs
for a waiting
time PDF, $\phi(t)\sim (t/\tau)^{-1-\alpha}$
or $\hat{\phi}(s) \sim 1 -(\tau s)^{\alpha}$,
where $0<\alpha<1$.
In the diffusion limit, the jump length PDF 
is given by
$\lambda (k)\sim 1-\sigma^2k^2/2$, 
where $\sigma^2$ is the second moment of the jump length PDF. 
In this case Eq.\ (\ref{eq:x2s2}) 
yields for $t\rightarrow \infty$
\begin{equation}
\left\langle \left\langle X^{2}(t)
\right\rangle \right\rangle=
\frac{\beta_{x}^{2}D_{x}}
{\sqrt{K_{\alpha}}\Gamma(2-\alpha/2)}
t^{1-\alpha/2},
\label{eq:sub}
\end{equation}
where $K_{\alpha}=\sigma^{2}/(2\tau^{\alpha})$ is
a generalized diffusion coefficient.
In other words, subdiffusion in the $y$-direction
with anomalous exponent $\alpha$ gives rise to
subdiffusive transport through the ramified
structure along the backbone with
exponent $1-\alpha/2$.
This result agrees with the result obtained considering a two-dimensional fractional diffusion equation to describe anomalous diffusion in the teeth and normal diffusion along the backbone (see \cite{MeIo13} for details).
Note that the transport process along the backbone and the teeth are very different. The transport along the backbone is always diffusive because the driving noise $\xi_{x} (t)$ is assumed white and Gaussian. However, the movement of particles along the teeth is governed by a waiting time PDF at a given point in the teeth. The anomalous exponent is $\alpha$ and for very long waiting time, that is $\alpha$ very small, the particles have a small probability of entering the teeth; it is far more likely that get swept along the backbone. Then, as $\alpha \rightarrow 0$, the probability of entering the teeth goes to zero and the transport along the comb is basically described by the transport along the backbone, i.e., it approaches to normal diffusion. On the other hand, as the motion in the teeth approaches
normal diffusive behavior, $\alpha \to 1$,
the MSD approaches the well-known behavior
$\left\langle \left\langle X^{2}(t)
\right\rangle \right\rangle\sim \sqrt{t}$
of simple random walks on combs
\cite{WeHa86,HabA87,Be06b}. If $\alpha$ governs both the motion along the backbone and the teeth as in \cite{MeIoCaHo15}, then the MSD scales as $t^{\alpha /2}$.

Analogously, to account for superdiffusion along the teeth
we consider an exponential waiting-time PDF
$\phi(t)=\exp(-t/\tau)/\tau$, i.e,
$\hat{\phi}(s) \sim 1 -\tau s$, and a heavy-tailed jump
length PDF, $\lambda(y)\sim \sigma^{\mu}\lvert y\rvert^{-1-\mu}$,
i.e., $\lambda (k)\sim 1-\sigma^{\mu}\lvert k\rvert^{\mu}$,
where $1<\mu<2$. In other words, the motion 
along the teeth, $Y(t)$,
is a L{\'e}vy flight. 
In this case,
Eq.\ (\ref{eq:x2s2}) yields
\begin{equation}
\left\langle \left\langle X^{2}(t)
\right\rangle \right\rangle=
\frac{2\beta_x^2D_x}{\mu 
K_{\mu}^{1/\mu}\sin (\pi/\mu)}
\frac{t^{1-1/\mu}}{\Gamma(2-1/\mu)},
\end{equation}
where $K_{\mu}=\sigma^{\mu}/\tau$
is
a generalized diffusion coefficient.
Interestingly, superdiffusive motion
in the $y$-direction also gives rise 
to subdiffusive transport through the ramified
structure, i.e., along the backbone, with the
anomalous exponent $1-1/\mu<1/2$.

For diffusive transport along the teeth,
$\lambda (k)\simeq 1-\sigma^2k^2/2$,
with a general waiting-time PDF $\phi(t)$,
we find after 
some algebra that Eq.\ (\ref{eq:x2s2}) reads
\begin{equation}
\left\langle \left\langle \hat{X}^{2}(s)
\right\rangle \right\rangle=
\frac{\sqrt{2}\beta_x^2D_x}{\sigma s^2} 
	\sqrt{\hat{\phi} (s)^{-1}-1}.
	\label{eq:msdn}
\end{equation}
If $\phi (t)$ has finite moments,
we expand the PDF for small $s$ to 
obtain $\hat{\phi} (s)^{-1}\simeq 1+\langle t \rangle s+\dotsb$. 
From Eq.\ (\ref{eq:msdn}) we recover the result 
$\left\langle \left\langle X^{2}(t)
\right\rangle \right\rangle \sim t^{1/2}$, 
regardless of the specific form of the waiting-time PDF. 
Finally, if we consider heavy-tailed PDFs 
for both the waiting times and the jumps lengths, i.e., 
$\hat{\phi} (s)=1-(\tau s)^{\alpha}$ and 
$\lambda (k)
\sim 1-\sigma^{\mu}\vert k\vert ^{\mu}$, 
we find from (\ref{eq:x2s2}) after some calculations 
$\left\langle \left\langle X^{2}(t)
\right\rangle \right\rangle \sim t^{1-\alpha/\mu}$, 
which predicts subdiffusive transport along the backbone.

\subsection{Fractional Brownian Motion and Fractal Time Process}

An interesting and well known non-standard random walks are the Fractional Brownian motion (FBM) and the Fractal Time Process (FTP). If the particles perform a FBM along the teeth, 
the diffusion equation reads
\begin{equation}
\frac{\partial P_{Y}(y,t)}{\partial t}=\alpha D_{\alpha}t^{\alpha-1}
\frac{\partial^{2} P_{Y}(y,t)}{\partial y^{2}},
\label{eq:FBMDE}
\end{equation}
where $0<\alpha<1$ and $D_{\alpha}$ is a generalized diffusion coefficient. 
The solution of Eq.\ \eqref{eq:FBMDE} is given by \cite{Lu01}
\begin{equation}
P_{Y}(y,t)=(4\pi Dt^\alpha)^{-1/2}\exp (-y^2/4Dt^\alpha).
\label{eq:FBMprop}
\end{equation}
Substituting this expression into Eq.\ \eqref{eq:r0} yields the following
expression for the MSD along the backbone,
\begin{equation}
\left\langle \left\langle X^{2}(t)
\right\rangle \right\rangle=
\frac{\beta_{x}^{2}D_{x}}
{\sqrt{\pi D_{\alpha}}(1-\alpha/2)}
t^{1-\alpha/2}.
\label{eq:MSDfBm}
\end{equation}
In other words, the transport along the backbone 
is subdiffusive with  exponent $1-\alpha/2$, 
as in the case of a subdiffusive CTRW, see Eq.\ (\ref{eq:sub}). 

If the motion of the particles along the 
teeth corresponds to the FTP, 
the diffusion equation reads
\begin{equation}
\frac{\partial P_{Y}(y,t)}{\partial t}=\frac{D_{\alpha}}{\Gamma(\alpha-1)}
\int_{0}^{t}\frac{dt'}{(t-t')^{2-\alpha}}
\frac{\partial^{2} P_{Y}(y,t')}{\partial y^{2}}.
\label{eq:ftp}
\end{equation}
The solution of Eq.\ \eqref{eq:ftp} in Laplace space 
is given by \cite{Lu01}
\begin{equation}
\hat{P}_{Y}(y,s)=\left[2\sqrt{D_{\alpha}}s^{1-\alpha/2}\right]^{-1}
\exp\left(-\lvert y\rvert s^{\alpha/2}/\sqrt{D_{\alpha}}\right).
\label{eq:ftpprop}
\end{equation}
Substituting this expression into Eq.\ \eqref{eq:x2sk} 
and taking the inverse Laplace transform, we find
\begin{equation}
\left\langle \left\langle X^{2}(t)
\right\rangle \right\rangle=
\frac{\beta_{x}^{2}D_{x}}
{\sqrt{D_{\alpha}}\Gamma(2-\alpha/2)}
t^{1-\alpha/2}.
\label{eq:MSDftp}
\end{equation}
In both cases, FBM and FTP the MSD  scales with time as for the case of 
a subdiffusive CTRW, see Eq.\ (\ref{eq:sub}).

\subsection{Fractal teeth}\label{sub:ft}
We next consider ramified structures
where the teeth consist
of branches with a spatial dimension different from one. 
In this Section we consider the case of 
particles undergoing a random walk 
on secondary branches with fractal structure.
The case of $n$-dimensional teeth will be studied
in Sec.\ \ref{sec:ntheeth}.
Equation \eqref{eq:r0} implies
that we only need to know the 
value of $P_{Y}(y=0,t)$.
Mosco \cite{Mo97} (see also Eq.\ (6.2) in Ref.\ \cite{MeFeHo10}) 
obtained the following expression for the propagator 
through a fractal in terms of the euclidean distance $r$, 
\begin{equation}
P_Y(r,t)\sim t^{-d_f/d_w}\exp \left[-c\left(\frac{r}{t^{1/d_w}}
\right)^{\frac{d_wd_{\min}}{d_w-d_{\min}}}\right],
\end{equation}
where $d_f$ and $d_w$ are the fractal and 
random walk dimensions, respectively,
and $d_{\min}$ corresponds to the fractal dimension 
of the shortest path between two given 
point in the fractal. 
Substituting $P_Y(r=0,t)\sim t^{-d_f/d_w}$
into Eq.\ (\ref{eq:r0}), we find for $t\to\infty$,
\begin{equation}
\langle \langle X^{2}(t)\rangle\rangle
\sim 
\begin{cases}
\ln(t), & d_f=d_w,\\[1ex]
t^{1-d_f/d_w}, & d_f<d_w,\\[1ex]
O(1), & d_f>d_w.
\end{cases}
\label{eq:ms2}
\end{equation} 
These results coincide with the scaling results 
predicted in \cite{FoBuCeVu13}.
If $d_f>d_w$, the MSD approaches a constant
value as time goes to infinity. This corresponds to stochastic
localization, i.e., transport failure \cite{DeHo00}.

\section{Dynamics in the teeth driven by external noise}\label{sec:teethLang}
We next consider that the motion
in the $y$-direction is given by
the Langevin equation \eqref{eq:le2}.
With the initial condition
$Y(0)=0$, Eq.\ (\ref{eq:le2}) yields:
\begin{equation}
Y(t)=\int_0^t\xi_y (t')dt'.
\label{eq:Y}
\end{equation}
Consequently, we can
express the PDF $P_{Y}(k,t)$ in terms
of the characteristic functional of the noise
$\xi_{y}(t)$, 
\begin{equation}
\Phi(k,t)=\left\langle \exp\left(ik
\int_{0}^{t}\xi_{y}(t')dt'\right)\right\rangle.
\label{15}
\end{equation}
Substituting Eq.\ (\ref{15}) into Eq.\ (\ref{eq:r0}),
we find
\begin{equation}
\left\langle \left\langle X^{2}(t)
\right\rangle \right\rangle=
\frac{ D_{x}\beta_{x}^{2}}{\pi}\int_{0}^{t}dt'
\int_{-\infty}^{\infty} \Phi(k,t') dk.
\label{eq:x2g}
\end{equation}
We have obtained a general expression for
the MSD of the transport through a ramified structure
for a given Langevin particle dynamics. 

\subsection{Colored Gaussian External Noise}

We assume that the particles move
along the teeth 
driven by a Gaussian colored noise
$\xi_{y}(t)$ with arbitrary 
autocorrelation 
$\left\langle\xi_{y}(t)\xi_{y}(t')\right\rangle=\gamma(t,t')$.
White noise corresponds
to the limiting case $\gamma(t,t')=\delta(t-t')$.
The characteristic
functional of a zero-mean Gaussian random process
is given by, see e.g. Ref.\ \cite{Kl05}, 
\begin{equation}
\Phi(k,t)=\exp\left[-k^{2}
\int_{0}^{t}dt'\int_{0}^{t'}\gamma(t',t'')dt''\right].
\end{equation}
We assume that the noise is stationary,
i.e., $\gamma(t,t')=\gamma(t-t')$. 
We change the order of integration 
and obtain
\begin{equation}
\Phi(k,t)=\exp\left[-k^{2}
\int_{0}^{t}dt'\gamma(t')(t-t')\right].
\end{equation}
Since
\begin{equation}
\mathcal{L}_t\left[ \int_{0}^{t}dt'\gamma(t')(t-t')\right]
= \frac{\hat{\gamma}(s)}{s^{2}},
\end{equation}
we can write the characteristic functional in the form
\begin{equation}
\Phi(k,t)=\exp\left[-k^{2}\mathcal{L}_{t}^{-1}
\left(\frac{\hat{\gamma}(s)}{s^{2}}\right)
\right],
\end{equation}
where $\mathcal{L}_{t}^{-1}$ denotes
the inverse Laplace transform.
Substituting this result into Eq.\ (\ref{eq:x2g}) 
and performing the integral
over $k$, we find
\begin{equation}
\left\langle \left\langle X^{2}(t)
\right\rangle \right\rangle
=\beta_{x}^{2}\frac{D_{x}}{\sqrt{\pi}}
\int_{0}^{t}dt'\left\{\mathcal{L}_{t'}^{-1}
\left[\frac{\hat{\gamma}(s)}{s^{2}}\right]\right\}^{-1/2}.
\label{eq:x2cc}
\end{equation}
Equation \eqref{eq:x2cc} is a 
concise relation between the MSD
of the transport along the backbone
and the statistical characteristics
of the stationary Gaussian noise 
driving
the motion along the teeth in 
term of its autocorrelation function $\gamma(t-t')$.
We define the noise intensity as 
$
D_{y}=(1/2)\int_{0}^{\infty}\gamma(t)dt
=\hat{\gamma}(s=0)/2,
$
according to Ref.\ \cite{HaJu95}.
If $D_y$
is finite and nonzero, 
the function $\hat{\gamma}(s)$
can be expanded in a power series expansion for 
small $s$. Up to the leading
order we find $\hat{\gamma}(s)/s^{2}
\simeq2D_{y}/s^{2}$, and 
$\mathcal{L}_{t'}^{-1}\left[\hat{\gamma}(s)/s^{2}\right]
\simeq2D_{y}t'$.
Therefore we obtain from Eq.\ (\ref{eq:x2cc}),
 \begin{equation}
\left\langle \left\langle X^{2}(t)
\right\rangle \right\rangle=
\frac{D_{x}\beta_{x}^{2}}
{2\sqrt{2D_{y}\pi}}t^{1/2}\quad\textrm{as}\quad t\to\infty,
\label{tint}
\end{equation}
i.e., 
the transport through the ramified structure
is subdiffusive with anomalous exponent $1/2$.

\begin{figure}[htbp]
\includegraphics[width=0.6\hsize]{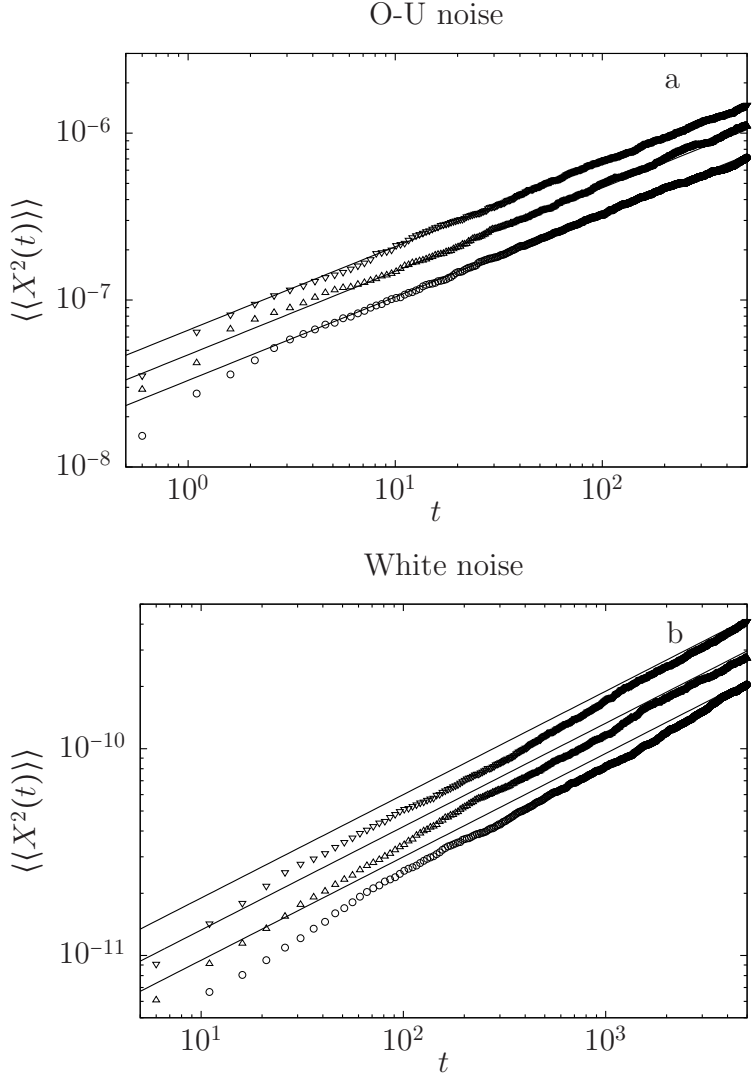}
\caption{MSD for different cases where the coupling functions 
has been taken as in Eq.\ (\ref{Cy}) with $\ell = 5\times 10^{-4}$. 
Panel a) Gaussian Ornstein-Uhlembeck noise, i.e., 
exponential correlation function $\gamma(t)=\sigma^2e^{-t/\tau}/2\tau$ 
and intensity $D_y=\sigma^2/4\tau^2$. The symbols represent 
numerical simulations for different values of the noise intensity;
circles: $D_y=1$, triangles: $D_y=0.25$, and  inverted triangles:
$D_y=1/16$. 
Panel b) Gaussian white noise with $D_y=1$ (circles), 
$D_y=0.5$ (triangles), and $D_y=1/4$ (inverted triangles). 
In both panels, $\beta_x =0.5$ and $D_x=1$. 
The straight solid lines correspond to 
the theoretical predictions given by Eq.\ (\ref{tint})}
\label{fig:f1}
\end{figure}

Figure 1 confirms the result provided 
by Eq.\ (\ref{tint}),
which implies that the MSD grows like $\sqrt{t}$ 
for long times for any Gaussian noise, 
regardless its correlation function.
We have shown that subdiffusive
transport with anomalous exponent 1/2
emerges under more general circumstances,
namely if the 
motion in the $x$-direction,
i.e., along the backbone,
is driven by \textit{any} white noise and
the motion along the teeth
is driven by any colored Gaussian 
noise with nonzero intensity.

\subsection{Non-Gaussian White External Noise}
We assume now that the particles move
along the teeth 
driven by non-Gaussian noise,
so-called L{\'e}vy noise. {\color{blue}{This noise in white in time, i.e., the autocorrelation function is $\left\langle\xi_{y}(t)\xi_{y}(t')\right\rangle=\delta(t-t')$.}}  Then $\xi_{y}(t)$ 
is 
the time derivative of a generalized
Wiener process $Y(t)$, i.e.,  
$Y(t)=\int_0^t\xi_y (t')dt'$, see Eq.\  (\ref{eq:Y}).
The 
random process $Y(t)$ has
stationary independent
increments on non-overlapping 
intervals \cite{DuSp05,DeHoHa09}. It
belongs to the class of L{\'e}vy processes,
and its PDF belongs
to the class of infinitely divisible distributions.
The characteristic functional of $Y(t)$ can be 
written in the form \cite{DuSp05}
\begin{equation}
\Phi(k,t)=\exp\left[t\int_{-\infty}^{\infty}dz
\rho(z)\frac{e^{ikz}-1
	-ik\sin(z)}{z^{2}}\right].
\label{eq:F2}
\end{equation}
Gaussian white noise corresponds to 
the kernel $\rho(z)=2\delta(z)$.
Symmetric L{\'e}vy-stable noise 
with index $\theta$ corresponds to the
power-law kernel 
$\rho(z)\sim\left\lvert z\right\rvert^{1-\theta}$ with $0<\theta<2$,
which yields
\begin{equation}
\Phi(k,t)
=\exp\left(-tD_{\theta}
\left\lvert k\right\rvert^{\theta}\right),
\label{carf}
\end{equation}
where $D_{\theta}$ is a generalized diffusion
coefficient. 
Substituting
this expression for $\Phi(k,t)$ into Eq.\
(\ref{eq:x2g}) we 
obtain
\begin{equation}
\langle \langle X^{2}(t)\rangle\rangle
\sim
\begin{cases}
\ln(t), & \theta=1,\\[1ex]
t^{1-1/\theta}, & 1<\theta\leq 2,\\[1ex]
O(1), & 0<\theta <1,
\end{cases}
\label{eq:ms1}
\end{equation}
as $t\to\infty$. 
If $\theta=2$, the characteristic functional (\ref{carf}) 
corresponds to the Gaussian one, 
and from (\ref{eq:ms1}) the MSD grows 
like $t^{1/2}$, as expected.
For $1<\theta<2$, the MSD displays
subdiffusive behavior,
and the anomalous exponent decreases
as $\theta$ decreases from 2 to 1. 
When it reaches the value $\theta=1$, Cauchy functional,
the MSD grows ultraslowly. This 
behavior has been observed before \cite{FoBuCeVu13,BoSo14},
but it appears here as a result of specific values 
of the characteristic parameters of the
noise that drives the motion along the teeth. 
Finally, if $0<\theta<1$,
the exponent is negative and the MSD 
approaches a constant value as time goes to infinity, 
i.e., stochastic
localization or transport failure occurs.

\begin{figure}[htbp]
\includegraphics[width=0.6\hsize]{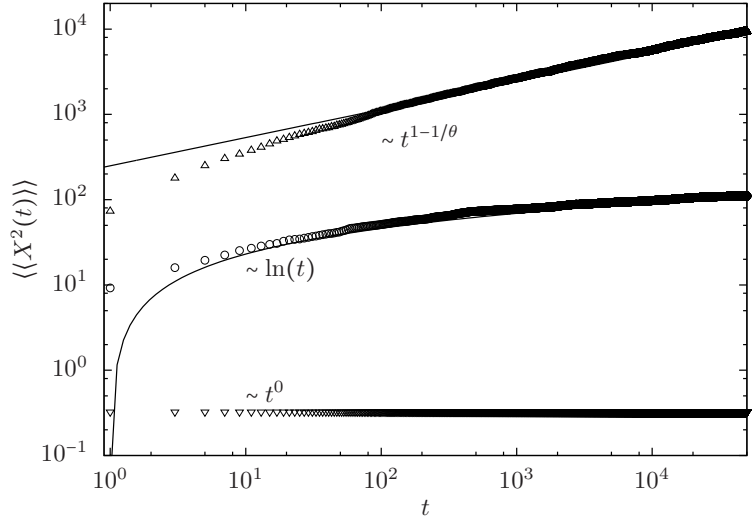}
\caption{MSD for three different values of the
exponent $\theta$. Monte Carlo simulations correspond 
to $\theta =1$ (circles), $\theta =1.5$ (triangles), 
and $\theta =0.5$ (inverted triangles). 
Solid lines correspond to the theoretical 
predictions given by Eq.\ (\ref{eq:ms1})}
\label{fig:f2}
\end{figure}

In Figure 2 we compare the analytical results 
provided by Eq.\ (\ref{eq:ms1}) 
with Monte Carlo simulations. 
Numerical and theoretical predictions show very good agreement 
for large $t$, 
where the results given by Eq.\ (\ref{eq:ms1}) hold. 

\subsection{Gaussian White Noise 
Along $n$-Dimensional Teeth}\label{sec:ntheeth}

Finally we consider a ramified structure consisting 
of a unidimensional backbone 
intersected by $n$-dimensional secondary 
branches at the same point $(x=ia,y_{l}=0)$, 
where $l=1,...,n$. To deal with the 
stochastic dynamics, we consider 
Eq.\ (\ref{eq:leq1}) together with the 
set of Langevin equations
\begin{equation}
\frac{dY_l}{dt}=\xi_{y_{l}}(t).
\label{25}
\end{equation}
Proceeding similarly as for the 
case $l=1$ and taking into account 
$\langle\xi_{x}(t)\xi_{x}(t')\rangle
=2D_{x}\delta (t-t')$, we find
\begin{equation}
\frac{dX^2}{dt}=2D_x\beta_x^2 
\prod_{l=1}^{n}[C_l(Y_l)]^2.
\end{equation}
Averaging over $Y_1,\dotsc,Y_n$ yields
\begin{equation}
\frac{d}{dt}\left\langle\left\langle 
X^2(t)\right\rangle \right\rangle=2D_x
\beta_x^2\prod_{l=1}^{n}\int_{-\infty}^{\infty}
(C_l[y_l])^2P_{Y_l}(y_l,t)dy_l.
\label{27}
\end{equation}
We assume again that the 
dynamics of $X(t)$ and $Y_{l}(t)$
are coupled within a narrow strip of
width $\ell$ around the
backbone, i.e., the coupling function 
$C_l[y_l]$ has the form given by Eq.\ \eqref{Cy}.

Integration of Eq.\ (\ref{27}) yields, in the limit 
$\ell \rightarrow 0$,
\begin{equation}
\left\langle\left\langle X^2(t)\right\rangle\right\rangle
=2D_x\beta_x^2\int _0^t 
\prod_{l=1}^{n}P_{Y_l}(0,t')dt'.
\end{equation} 
As in Eq.\ (\ref{15}), 
$P_{Y_l}(k_{l},t)=\langle \exp[ik_{l}Y_l(t)]\rangle$.
Integrating Eq.\ (\ref{25}),
we find the characteristic 
functional for each $\xi_{y_l}$,
\begin{equation}
\Phi(k_l,t)=\left\langle\exp 
\left( ik_l \int_0^t 
\xi_{y_l}(t')dt'\right)\right\rangle,
\end{equation}
and Eq.\ \eqref{eq:x2g} now reads
\begin{equation}
\left\langle\left\langle X^2(t)\right\rangle \right\rangle
=\frac{2D_x\beta_x^2}{(2\pi)^{n}}
\int_0^{t} dt'
\prod_{l=1}^{n}\int_{-\infty}^{\infty}dk_{l}
\Phi(k_l,t').
\label{29}
\end{equation}
We consider the case that
the $\xi_{y_l}(t)$ are uncorrelated 
Gaussian white noises, i.e., 
$\langle\xi_{x_m}(t)\xi_{x_l}(t')\rangle
=2D_{y_m}\delta_{ml}\delta (t-t')$,
where $m,l=1,\dotsc,n$. Their characteristic 
functional  is $\Phi(k_l,t)=\exp(-tD_{y_l}k_{l}^{2})$. 
Substituting this result into Eq.\ (\ref{29}), we find
\begin{equation}
\left\langle\left\langle X^2(t)\right\rangle\right\rangle
\sim
\begin{cases}
t^{1/2}, & n=1,\\[1ex]
\ln(t), & n=2,\\[1ex]
O(1), & n>2.
\end{cases}
\label{eq:msf}
\end{equation}
Note that the transport shows behavior similar to that
of a comb with fractal teeth, see Sec.\ \ref{sub:ft}.

\section{Secondary branches with finite length}\label{sec:teethfin} 
If the range $\mathcal{D}$
of $Y(t)$ corresponds to a 
finite interval, it is convenient
to work directly with Eq.\ (\ref{eq:r0}), 
particularly if the dynamics on 
the secondary branches is described 
by a diffusion equation.
As an example consider
the case of normal 
diffusion described by the equation 
$\partial_t P_Y=D_y\partial_{yy}P_Y$ 
along one-dimensional branches in 
the $y$-direction of length $2L$ 
with reflecting boundary conditions, 
$(\partial_y P_Y)_{y=\pm L}=0$,  
and initial condition $P_Y(y,0)=\delta (y)$. 
The solution $P_{Y}(y,t)$ is given
by the Fourier series expansion
\begin{equation}
P_Y(y,t)=\frac{1}{2L}+\frac{1}{L}\sum_{n=1}^{\infty}
\exp\left(-\frac{n^2\pi^2D_y}{L^{2}}t\right)
\cos\left(\frac{n\pi y}{L}\right).
\label{eq:pex}
\end{equation}
By inserting (\ref{eq:pex}) into (\ref{eq:r0}) we find after some algebra
\begin{equation}
\langle \langle X^{2}(t)\rangle\rangle
=\frac{\beta_x^2 D_x}{L}t +\frac{\beta_x^2 D_x L}{3D_y}-\frac{2\beta_x^2 D_x L}{\pi^2 D_y}\sum _{n=1}^{\infty}n^{-2}\exp\left(-\frac{n^2\pi^2D_y}{L^{2}}t\right)
\end{equation} 
Consequently in the limit $t\to\infty$

\begin{equation}
\langle \langle X^{2}(t)\rangle\rangle
\simeq\frac{\beta_x^2 D_x}{L}t,
\label{msdn}
\end{equation}
i.e., transport through
the comb is normal diffusion as expected. 

We compare this result 
with the case where the diffusion 
along the teeth is anomalous. 
The equation for subdiffusion 
along one-dimensional branches in 
the $y$-direction of length $2L$ is 
given by the fractional diffusion equation
$
\partial _t P_Y= {}_{0}
\mathcal{D}_{t}^{1-\alpha}
K_{\alpha}\partial_{y}^2P_{Y}, 
$
where ${}_{0}\mathcal{D}_{t}^{-\alpha}$ 
is the Riemann-Liouville fractional derivative 
with $0<\alpha<1$ \cite{MeKl00} and $K_{\alpha}$ is
a generalized diffusion coefficient. 
The solution $P_{Y}(y,t)$ is given by
\begin{equation}
P_Y(y,t)=\frac{1}{2L}+\frac{1}{L}\sum_{n=1}^{\infty}
E_{\alpha}\left(-\frac{n^2\pi^2K_{\alpha}}
{L^2}t^{\alpha}\right)\cos\left(\frac{n\pi y}{L}\right),
\label{sol1}
\end{equation}
where $E_{\alpha}(z)$ is the
Mittag-Leffler function.
Using Eqs.\ (\ref{eq:r0}) and 
(\ref{sol1}), $E_{\alpha}(z)=E_{\alpha,1}(z)$, and
the integration formula \cite{Po99}
\begin{equation}
\int_{0}^{t}d\tau E_{\alpha,\beta}\left(\lambda \tau^{\alpha}\right)
\tau^{\beta-1}=t^{\beta}E_{\alpha,\beta+1}\left(\lambda t^{\alpha}\right)
\label{eq:MLint},
\end{equation}
we obtain the MSD
\begin{equation}
  \langle \langle X^{2}(t)\rangle\rangle
= \frac{\beta_x^2D_xt}{L}+ \frac{2\beta_x^2D_xt}{L}\sum_{n=1}^{\infty}
E_{\alpha,2}\left(-\frac{n^2\pi^2K_{\alpha}
	t^{\alpha}}{L^2}\right),
\end{equation}
where $E_{\alpha,\beta}(z)$
is the Generalized Mittag-Leffler function.
The long-time behavior of the Mittag-Leffler function is
given by \cite{Ba53}
\begin{equation}
E_{\alpha,2}\left(-\frac{n^2\pi^2K_{\alpha}
	t^{\alpha}}{L^2}\right)\sim \frac{L^2}
	{\Gamma (2-\alpha)n^2\pi^2K_{\alpha}
	t^{\alpha} },
	\label{eq:MLasymp}
\end{equation}
and the MSD reads
\begin{equation}
  \langle \langle X^{2}(t)\rangle\rangle
  = \frac{\beta_x^2D_x}{L}t+\frac{\beta_x^2D_x L}
  {3\Gamma (2-\alpha)K_{\alpha}}t^{1-\alpha},
  \label{msdf}
\end{equation}
where we have used $\sum_{n=1}^{\infty}1/n^{2}=\pi^{2}/6$.
It is clear that for $t\rightarrow \infty$ the first term 
of the right hand side of (\ref{msdf}) is dominant and the MSD displays normal  
diffusive behavior. 

Having studied the effect of subdiffusion in finite-length
teeth, we
now consider the case where particles perform 
superdiffusive motion in the teeth. 
The equation for $P_Y(y,t)$ is given by 
$\partial_t P_Y=D_{\mu}\partial_{y}^{\mu}P_Y$
with $1 < \mu <2$ and
with the same boundary and initial conditions as in the previous cases.
Superdiffusion is described by the fractional 
derivative $\partial_{y}^{\mu}$, 
which corresponds to a heavy-tailed jump length PDF, 
and $D_{\mu}$ is a generalized transport
coefficient. 
The eigenvalue problem $\partial_{y}^{\mu}\psi_n(y)=e_n\psi_n(y)$
has been considered in \cite{Io15}. The L{\'e}vy operator in a
box of size $2L$ reads
\begin{equation}\label{prt-beta}
\partial_{y}^{\mu}f(y)=\int_{-L}^{L}\left[\frac{1}{2\pi}
\int_{-\infty}^{\infty}
(-\lvert k\rvert^{\mu})e^{-ik(y-y')}dk\right]f(y')dy'.
\end{equation}
As it follows from \cite{Io15}, the eigenfunctions are $\psi_n(y)=
\cos(n\pi y/L)$, where $ n=0,1,2,\dotsc$,
with corresponding eigenvalues $e_n=-(\pi n/L)^{\mu}$.
The PDF in the teeth reads now 
\begin{equation}
P_Y(y,t)=\frac{1}{2L}+\frac{1}{L}\sum_{n=1}^{\infty}
\exp\left(-\frac{n^{\mu}\pi^{\mu}D_{\mu}}{L^{\mu}}t\right)
\cos\left(\frac{n\pi y}{L}\right).
\label{eq:pex-a}
\end{equation}
Following the same steps to obtain (\ref{msdn}) from (\ref{eq:pex}) we find here  the asymptotic result 
\begin{equation}\label{sol-last-a}
\langle \langle X^{2}(t)\rangle\rangle
=\frac{\beta_x^2 D_{x}}{L}t
\quad\text{as}\quad t\to\infty.
\end{equation}
We have shown that the transport 
along the backbone is diffusive for finite
length teeth, if the transport regime of the particles in the teeth
is normal diffusion, subdiffusion, and superdiffusion.

The robustness of the diffusive behavior of the MSD
along the backbone can be understood as follows.
If the random motion of the particles along the
finite teeth with reflecting boundary
conditions is homogeneous and unbiased, then
$P_{Y}(y,t)\to 1/(2L)$ as $t\to \infty$.  The function
system
\begin{equation}
\frac{1}{\sqrt{2L}},\, 
\frac{1}{\sqrt{L}}\cos\left(\frac{n\pi y}{L}\right),\quad
n=1,2,\dotsc,
\end{equation}
is a complete orthonormal system on $[-L,L]$. Consequently, the
PDF of the particle motion along the teeth, with initial 
condition $P_Y(y,0)=\delta (y)$,
can be written as
\begin{equation}
P_Y(y,t)=\frac{1}{2L}+\frac{1}{L}\sum_{n=1}^{\infty}
T_{n}(t)
\cos\left(\frac{n\pi y}{L}\right),
\label{eq:finite}
\end{equation}
with $T_{n}(0)=1$ and $T_{n}(t)\to 0$ as $t\to \infty$.
If $T(t)\equiv \sum_{n=1}^{\infty}
T_{n}(t)$ is well-defined, i.e., the series converges,
for $t$ sufficiently large and if there exists a constant $C$
with $0\leq C <\infty$, such that $(1/t)\int_{0}^{t}dt' T(t')\to C$
as $t\to\infty$, then
the MSD displays again normal diffusive behavior. These conditions
are satisfied for the three cases analyzed above. In other words, if the teeth are finite , then the reflecting boundary conditions will give rise to a uniform distribution along the teeth for all types of transport. That is, the nature of
the transport, anomalous or not, plays no role. 
This is due to a balance reached between particles within the teeth and those in the backbone. Although subdiffusive transport in the teeth makes that mean residence times within the teeth can diverge, this is balanced by the fact that typical times of departure from the backbone also diverge asymptotically with the same anomalous exponent. So, both effects compensate to keep $P_{Y}(y=0,t)$ constant asymptotically for large times, so the MSD will grow linearly in time according to Eq. (\ref{eq:r0}). In the Appendix we provide a more formal justification of this idea by studying the asymptotic behavior of $P_{Y}(y=0,t)$ as a function of the backbone-teeth time dynamics.
Therefore, since the
transport along the backbone itself is diffusive, being driven by white noise,
we expect to obtain a diffusive scaling for the MSD.

\section{Conclusions}\label{sec:concl}
We have adopted a general Langevin formalism to explore
transport through ramified comb-like structures.
The transport through the structure is characterized
by the behavior of the MSD along the backbone.
We have derived an exact analytical expression, given 
in Eqs.\ \eqref{eq:r0} -- \eqref{eq:x2s}, that
allows us to determine the MSD explicitly from the 
PDF of the motion along the secondary branches, $P_Y(y,t)$, 
i.e., the probability of a particle to be at point $y$ 
of a secondary branch at time $t$.

If the secondary branches 
have finite length and reflecting boundary
conditions, then under some mild conditions the transport
regime along the teeth does not matter and
the MSD is proportional to $t$, 
indicating standard diffusion.
We have shown this explicitly for diffusive, subdiffusive, 
and superdiffusive motion
along the secondary branches.
If the secondary branches have infinite length, then
both subdiffusion and superdiffusion along the teeth
generate a subdiffusive MSD along the backbone. 
Therefore, the finite or infinite length 
of the secondary branches plays a crucial 
role for the transport along the overall structure. 
 
Another interesting situation arises if the 
dynamics of the particles along the secondary branches 
are described directly by a Langevin equation.
For this case we have obtained an 
exact analytical formula, see Eq.\ (\ref{eq:x2g}),
that relates the MSD along the backbone 
to the characteristic functional of the 
noise $\xi_y (t)$ driving the motion 
along the secondary branches. 
This expression is completely general
and holds  
for any noise $\xi_y (t)$.
We have considered several different situations. 
For Gaussian colored noise $\xi_y (t)$, we have shown 
that if the noise intensity is finite and nonzero, 
then the MSD grows like $t^{1/2}$ along the backbone. 
We have checked this result with Monte Carlo simulations, 
performed for the case of Gaussian white noise 
and exponentially correlated Gaussian noise, 
i.e., Ornstein-Uhlenbeck noise. In addition, 
we have also considered that $\xi_y (t)$ 
is white but non-Gaussian noise. In this case 
our interest has been focused on  
symmetric L{\'e}vy-stable noise with 
exponent $\theta$. We have found that 
the MSD along the backbone grows ultraslowly 
like $\ln (t)$,
if the PDF of the white noise $\xi_y (t)$ 
is a Cauchy distribution, $\theta =1$. 
For $0<\theta<1$, the MSD exhibits 
stochastic localization, i.e., it approaches 
asymptotically a constant value,
while for $1<\theta<2$ the MSD exhibits subdiffusion. 
Excellent agreement is found with Monte Carlo simulations.
We have also considered multidimensional and fractal
secondary branches. We have obtained 
different behaviors like ultraslow motion, 
subdiffusion, and stochastic localization in terms 
of the dimension of the secondary branches. 

In summary, we have shown in this work 
how particles moving through a simple regular 
structure, namely a comb, are able to 
display a variety of 
macroscopic transport regimes, namely transport failure 
(stochastic localization), 
subdiffusion, or ultraslow diffusion,  
depending on whether the 
secondary branches have finite or infinite length but also 
on the statistical properties of the noise
that drives the motion along them. 
We expect our results to find applications to 
the description of the movement of organisms and animals 
through ramified structures like river networks,
ecological corridors, etc.        

\section*{Appendix}

In Section VI we have seen that diffusive properties in the backbone do not change qualitatively by introducing different modes of transport (superdiffusive, subdiffusive) within the teeth. Intuitively, one expects that the transport properties in the backbone are mainly determined by the dynamics of entrance within the teeth and return from it (since only particles at $y=0$ contribute to the transport in the backbone).

To clarify this connection, we here derive the dependence of $P_{Y}(y=0,t)$ (which determines the mean square displacement through Eq. (\ref{eq:r0})) with the typical times the particle stays in the teeth.
We introduce $\psi_1(t)$ as the probability distribution of times a particle stays in the backbone before entering into the teeth, and $\psi_2(t)$ as the corresponding distribution of times the particle spends within the teeth before returning to the backbone. So, the mean value of $\psi_2(t)$ determines the mean residence time within the teeth. The probability that a particle is at $y=0$ at an arbitrary time $t$ will be then given by
\begin{equation}
P_{Y}(y=0,t)= \Psi_1(t) + \psi_1(t) * \psi_2(t) * \Psi_1(t) + \psi_1(t) * \psi_2(t) * \psi_1(t) * \psi_2(t) * \Psi_1(t) + \ldots 
\end{equation}
where $\Psi_1(t)$ is the \textit{survival} probability of $\psi_1(t)$, i.e. $\Psi_1(t) = \int_{t}^{\infty} \psi_1(t')dt'$, and the asterisk denotes time convolution; also, we have here implicitly assumed that at $t=0$ all the particles are located in the backbone, $P_{Y}(y=0,0)=1$. In the previous expression, the first term on the rhs represents those particles which have not left yet the backbone at time $t$, the second term corresponds to those that are currently at the backbone after a previous excursion within the teeth, the third term represents those particles that have performed two previous excursions within the teeth, and so on.

Using Laplace transform to deal easily with the time convolution operators, we find
\begin{equation}
\hat{P_{Y}}(y=0,s) = \frac{\hat{\Psi_1} (s)}{1- \hat{\psi_1} (s) \hat{\psi_2} (s)}
\label{generic}
\end{equation}
where the hat denotes the Laplace transform, and $s$ is the Laplace argument.

Now that we have reached a generic expression connecting the backbone-teeth time dynamics to $\hat{P}_{Y}(y=0,s)$, we can study how this expression behaves in the long-time (or equivalently, small $s$) regime. For this, we assume that the distributions of times within the backbone and within the teeth follow generic \textit{anomalous} scaling in the asymptotic regime through $\psi_1(t) \sim t^{-1-\alpha_1}$ and $\psi_2(t) \sim t^{-1-\alpha_2}$, for $t \rightarrow \infty$. With the help of Tauberian theorems we can translate this to Laplace space and obtain finally from (\ref{generic})
\begin{equation}
\lim_{s \rightarrow 0} \hat{P_{Y}}(y=0,s) \sim s^{\alpha_1 -1 -\min (\alpha_1, \alpha_2)}
\label{exponent}
\end{equation}

This expression confirms our results above in Section VI. If the \textit{anomalous} exponent determining the entrance within the teeth and the return from it satisfies $\alpha_1 \leq\alpha_2$ then we get $\lim_{s \rightarrow 0} \hat{P}_{Y}(y=0,s)\sim s^{-1}$, or equivalently $\lim_{t \rightarrow \infty} P_{Y}(y=0,t)\sim \mathrm{const}$, and then the transport in the backbone is always diffusive independent of $\alpha_i$ with $i=1,2$. This will be the case for normal diffusion within the teeth, and also for \textit{anomalous} transport within the teeth determined by power-law asymptotic decay of waiting times (see, e.g. \cite{metzler2004}, for details and a deeper discussion on this point). Additionally, we observe from (\ref{exponent}) that only in the case of an imbalance in the backbone-teeth dynamics (so $\alpha_1 > \alpha_2$) we could obtain a different (non-diffusive) result.

\section*{Acknowledgments}
This research has been partially supported by 
Grants No. CGL2016-78156-C2-2-R by the Ministerio de 
Econom\'{\i}a y Competitividad  and by SGR 2013-00923 by the 
Generalitat de Catalunya.  AI was also supported by the 
Israel Science Foundation (Grant No.
ISF-1028). VM also thanks the University 
of California San Diego where part of this 
work has been done.

\bibliography{langcomb}

\end{document}